Angela Vasanelli[1], Yanko Todorov[1], Baptiste Dailly[2], Sebastien Cosme[2], Djamal Gacemi[1], Andrew Haky[1], Isabelle Sagnes[3], and Carlo Sirtori[1]

1. Laboratoire de physique de l'Ecole Normale Supérieure, ENS, Université PSL, CNRS, Sorbonne Université, Université de Paris, Paris, France
2. Laboratoire Matériaux et Phénomènes Quantiques, CNRS, Université de Paris, Paris, France
3. Centre for Nanosciences and Nanotechnology, CNRS, Universite Paris-Saclay, UMR 9001, 10 Boulevard Thomas Gobert, 91120, Palaiseau, France

**Semiconductor quantum plasmons for high frequency thermal emission**

**Abstract:** Plasmons in heavily doped semiconductor layers are optically active excitations with sharp resonances in the 5-15 μm wavelength region set by the doping level and the effective mass. Here we demonstrate that volume plasmons can form in doped layers of widths of hundreds of nanometers, without the need of potential barrier for electronic confinement. Their strong interaction with light makes them perfect absorber and therefore suitable for incandescent emission. Moreover, by injecting microwave current in the doped layer, we can modulate the temperature of the electron gas. We have fabricated devices for high frequency thermal emission and measured incandescent emission up to 50MHz, limited by the cutoff of our detector. The frequency dependent thermal emission is very well reproduced by our theoretical model that let us envision a frequency cutoff in the tens of GHz.

# 1 Introduction

Plasmons are collective excitations of electron gas that concentrate the interaction with light. In metals they have been studied for more than a century as volume and surface plasmons and have sparked fundamental semiclassical theories that we are still using to determine the high frequency conductivity and the upper limit of transparency (the plasma frequency) of many metals and highly doped semiconductors. The Drude theory, revisited then by Sommerfeld is the most successful story. Originally conceived to describe the thermal properties of metals, it contains also the essential features on how light couples to an ensemble of electrons and is still largely used for explaining the dielectric constant of conductive solids.

Since the 70's plasmons have also been studied in very narrow metallic films of few tens of Angstroms, where size confinement gives rise to sizable shifts of their energy[1–3]. Moreover, at the edges of the potential well plasmons develop a huge optically active dipole and become easily observable by spectroscopy techniques[4]. The resulting optical resonance is called Berreman or Ferrell-Berreman mode. The advent of nanotechnology has brought this field much farther by making available extremely small metallic or semiconductor particles in which plasmonic optical resonances can be tuned by size confinement[5,6]. Remarkably, plasmonics with objects of nanometric dimensions has permitted to concentrate the electromagnetic radiation well below the diffraction limit and has open the field of nano-antenna[7,8]. The study and the understanding of these properties are of major interest today and go under the name of quantum plasmonics[9].

Another challenge that is actively studied in this field is the generation of infrared radiation by thermally excited plasmons[10–12]. Due to their superradiant nature, excited plasmons radiatively decay by spontaneous emission[13]. Thermal radiation is of major importance for several applications, such as lighting, energy harvesting and management, sensing, tagging and imaging[14]. There are several ways of engineering thermal emission by exploiting a material resonance, such as phonon modes[15–17], optical transitions in a quantum well[18], or plasmonic resonances[19]. The interest of using plasmons is that they have huge dipoles and produce sharp optical resonances that can be as dark as a perfect blackbody on a narrow frequency band, ideal for a thermal emitter. Moreover, in a semiconductor the energy of the plasmonic resonance can be adjusted by selecting the level of doping. Dynamic control of incandescence[20] can be achieved through time modulation of the electronic temperature. This can be obtained by transferring energy through electrical injection either in the electron gas or in an adjacent resistance. The combination of electrical injection and the very low thermal capacitance of the electron gas allows a high frequency modulation of the temperature and therefore of the thermal emission.

In this work we show that mid-infrared optically active plasmons in a highly doped GaInAs layers embedded between two sharp AlInAs barriers are substantially identical to those arising in highly doped GaInAs layers without the confining barrier. In the following of the paper we will concentrate on the realisation of incandescent devices in which the emission spectrum is controlled by a Berreman mode. The modulation of the current injected in the doped layer induces fast temperature variations of the electron gas. In this configuration, we were able to modulate the thermal emission up to 50 MHz, the highest frequency response of our detector.

## 2 Volume plasmons in highly doped semiconductor layers

Bulk plasmons are collective excitations of a three dimensional electron gas existing in metals and semiconductors. In thin films with thickness smaller than their wavelength, these excitations develop dipole at the edges, due to instantaneous charge separation, and they can therefore emit or absorb photons from free space radiation[1,2,21]. These optically active confined plasmon modes are called Berreman modes. In highly doped semiconductor layers their optical resonance, at the plasma frequency $\Omega_P$, is in the mid-infrared range that correspond to optical wavelengths between 5 µm and 10 µm, depending on their effective mass and doping:

$$\Omega_p = \sqrt{\frac{e^2}{\epsilon_0 \epsilon_s} \frac{N_v}{m^*}} \qquad (1)$$

with $e$ the electronic charge, $\epsilon_0$ the vacuum electric constant, $\epsilon_s$ the background dielectric constant, and $m^*$ the effective mass.

Fig. 1a illustrates the plasmons that we are investigating. They develop an optical dipole in the direction perpendicular to the layer and they are spatially modulated in the plane, thus possess a wavevector, $k_{//}$. This quantity is the quantum number that is preserved in emission or absorption and can be selected by changing the angle of emission/detection, $\theta$, measured from the normal to the surface.

Plasmons appear in layers embedded between two confining barriers in a wide well configuration (Fig. 1b). The barriers however are not necessary as a simple heavily doped layer in an original "flat potential" bends the conduction band profile due to charge separation and form a potential well (Fig. 1c). In both cases, the confining potential forms a set of discrete electronic energy levels which are the basis that we use for the theoretical description of these collective excitations. Given the wide dimension of the potential wells, 150nm, it is remarkable that the electronic wavefunctions maintain their coherence and still form a set of energy states separated of few meV. The role of electronic confinement on plasmonic resonances with different potential geometries is therefore crucial to bring quantum engineering to the realm of plasmonics[22]. Moreover, a semiconductor platform instead of metals brings clear evidence on the quantum nature of the electrons building the plasmonic response. Fig. 1b and 1c show measured normalized absorption spectra from two 150nm doped layers that correspond to the potentials sketched in the figure. In both cases the absorption has a Lorentzian shape and peaks at ~ 160 meV, in excellent agreement with our theory reported in ref. [23]. Surprisingly, in spite of the disorder introduced by the donors, the spectra measured at small angles are characterized by sharp resonances with a quality factor on the order of 20. At higher angles the collective excitations are dominated by spontaneous superradiant emission[13] and become radiatively broadened (Fig. 2).

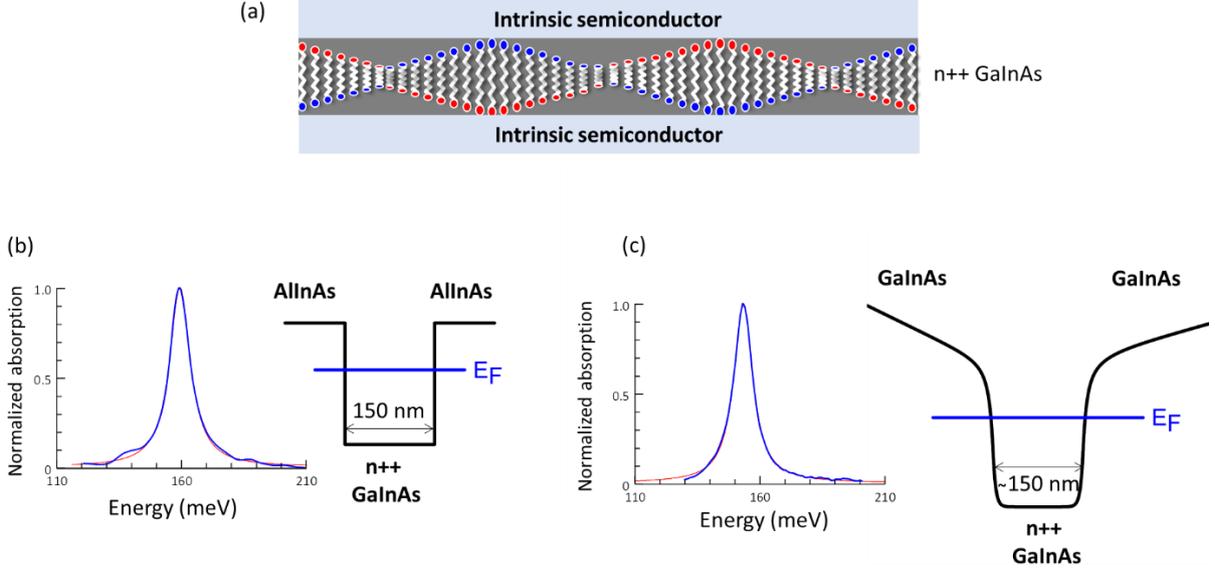

*Figure 1: (a) Sketch of the system: a highly doped GaInAs layer embedded between two intrinsic semiconductor layers sustains a Berreman mode characterized by a dipole along the growth direction. (b) and (c) Conduction band profile and normalized absorption spectrum measured at 10° internal angle (blue line) for a sample in which the two intrinsic layers are AlInAs barriers (panel b) or intrinsic GaInAs layers (c). Red lines are Lorentzian fit of the absorption spectra.*

## 3  Superradiance, perfect absorption

The very strong interaction that plasmons have with light is intimately related to their collective nature, i.e. they are a symmetric superposition of all the optically active electronic transitions of the system. The optical properties of the plasmon mode have been calculated by solving quantum Langevin equations in the input-output formalism[24]. The bright plasmon is coupled with two bosonic baths: a bath of electronic excitations and a bath of free space photons. The coupling with the electronic bath is considered phenomenologically, and it gives rise to a non-radiative broadening $\gamma$ of the plasmon mode, which is typically 20 times smaller than the plasmon frequency $\Omega_P$. The coupling between the plasmon and the free space photons is calculated thanks to our quantum model. It results in a radiative decay rate $\Gamma$, which is a function of the angle of emission, $\theta$, and of the photon frequency $\omega$ through:

$$\Gamma(\theta, \omega) = \Gamma_0 \frac{\omega}{\Omega_P} \frac{\sin^2 \theta}{\cos \theta} \qquad (2)$$

$\Gamma_0$ depends on the oscillator strength of the plasmon and its complete expression can be found in reference[24]. It can be shown that $\Gamma_0$ is proportional to the density of the electron gas, an indication of the superradiant nature of the plasmon emission[13]. The $\sin^2 \theta$ dependence in eq. (2) comes from the fact that the plasmon has a dipole oriented along the growth direction, while the $1/\cos \theta$ dependence is associated with the photonic density of states. The angular dependence of the radiative rate allows us to vary the interaction between the plasmon and the electromagnetic field by simply changing the light-propagation angle. As it will be discussed later, this permits to experimentally investigate different regimes of the light-matter interaction. It is also important to note that in our quantum model the spontaneous emission of the plasmons into the free space is calculated without employing a perturbative approach, and it results in a dependence of Γ on the photon frequency which is characteristics of non-Markovian dynamics.

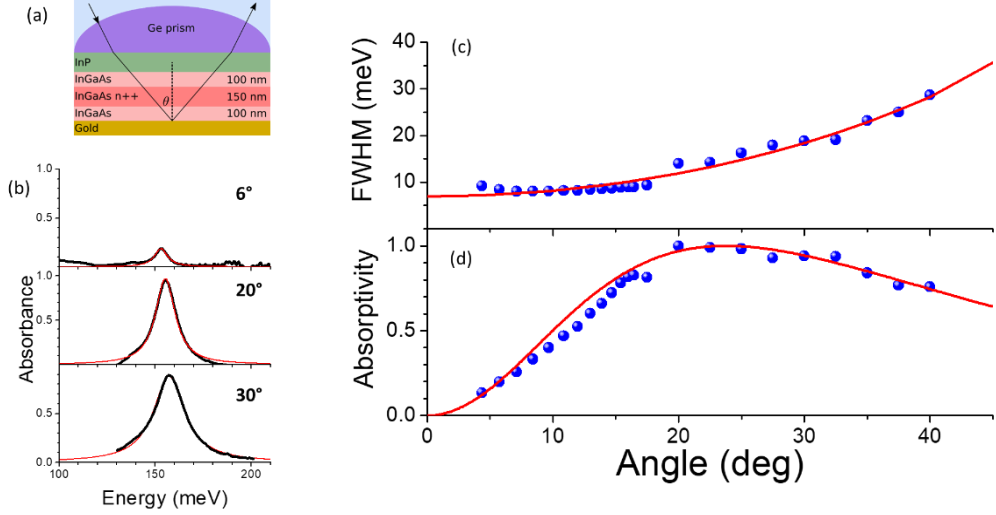

*Figure 2: (a) Sketch of the sample and of the geometry used for reflectivity experiments. (b) Absorption spectra at three different internal angles (black lines) and Lorentzian fit of the resonances (red lines). (c) Full width at half the maximum (FWHM) of the absorption peak (blue symbols) as a function of the internal angle. Red line presents the calculated broadening, including radiative and non-radiative contribution. (d) Peak absorptivity extracted from the absorption spectra as a function of the angle, compared with the calculated one (red line) by using eq. 3.*

The angular behavior of the plasmon spectra from the sample without the confining barriers (Fig. 1c) has been experimentally investigated by performing reflectivity spectra, after evaporating a gold mirror on the surface. In order to span a wide angular range, we have illuminated the sample through a Germanium hemispherical lens, as sketched in Fig. 2a. In this way absorption spectra at internal angles between 6° and 40° have been measured. Few representative spectra are presented in Fig. 2b, showing the Berreman mode and its evolution with the angle in terms of peak absorptivity and linewidth. In Fig. 2c the full width at half maximum (FWHM) of the absorption peaks, $\Gamma_{tot}$ is plotted as a function of the angle, θ. The results are well reproduced by equation (2) assuming that $\Gamma_{tot} = \gamma + \Gamma(\theta, \Omega_P)$ is the sum of the radiative and nonradiative contribution. By changing the propagation angle of the light, the radiative rate, Γ, varies from 0 to a value much bigger than the non-radiative rate γ, as it can be seen in Fig. 2c. At 6° internal angle, the radiative contribution to the plasmon linewidth is negligible, and therefore we can extract the value of the non-radiative broadening, γ = 8 meV. The quality factor of the plasmon mode, centered at 153 meV, is thus 19. This value is quite remarkable as the plasmon is simply defined by the presence of a highly doped GaInAs region between two intrinsic GaInAs layers.

Figure 2d presents the peak absorptivity as a function of the internal angle extracted from the measured absorptivity spectra (symbols) and calculated (solid line) by using the result of our model which is expressed below, in equation (3). As expected from our quantum model, the peak absorptivity increases with increasing the angle, until a critical coupling angle $\theta_c = 20°$ is reached. For greater angles, the peak absorptivity decreases while the linewidth increases. We also observe a blue shift of the plasmon peak, which can be well described as a cooperative Lamb shift[25]. Notice that the perfect absorption gives rise to a perfect blackbody emitter at the frequency of the Berreman mode.

The plasmon absorptivity is calculated by considering an optical input in our quantum model. In the presence of a single bright plasmon mode, close to a perfect gold mirror, the absorptivity is given by the following formula[24]:

$$\alpha(\theta,\omega) = \frac{\frac{4\Omega_P^2}{(\Omega_P+\omega)^2}\gamma\Gamma(\theta,\omega)}{(\omega-\Omega_P)^2 + \frac{4\Omega_P^2}{(\Omega_P+\omega)^2}\left(\frac{\gamma}{2}+\frac{\Gamma(\theta,\omega)}{2}\right)^2} \quad (3)$$

This expression is obtained beyond the rotating wave approximation and it takes into account the anti-resonant terms of the light matter interaction, responsible of the ultra-strong light-matter coupling[26]. Our model perfectly reproduces the experimental results, without using any fitting parameter, as it can be seen in Fig. 2d. Three different regimes of the light-matter interaction can be identified. For small angles, $\gamma \gg \Gamma(\theta,\omega)$, the absorptivity takes the following form for frequencies close to the peak:

$$\alpha(\theta,\omega) \approx \frac{2}{\gamma}\frac{2\Gamma}{1+\left(\frac{\omega-\Omega_P}{\gamma/2}\right)^2} \quad (4)$$

The absorption spectrum is thus Lorentzian, with a full width at half the maximum given by the non-radiative broadening. The amplitude of the peak is proportional to twice the radiative rate $\Gamma \approx \Gamma_0 \frac{\sin^2\theta}{\cos\theta} \approx \Gamma_0 \theta^2$ due to the presence of the gold mirror. This is exactly the result expected in a perturbative description of the plasmon absorption. In this regime the absorptivity peak increases with the angle, while the spectrum progressively broadens.

The second regime occurs at the angle $\theta_c$ for which the radiative rate equals the non-radiative one: $\gamma = \Gamma(\theta_c, \Omega_P)$. In this case the absorption spectrum is still a Lorentzian function, with unitary peak absorptivity and with full width at half the maximum $2\gamma$:

$$\alpha(\theta_c,\omega) \approx \frac{1}{1+\left(\frac{\omega-\Omega_P}{\gamma}\right)^2} \quad (5)$$

This is the critical coupling regime, giving rise to perfect absorption at the plasmon frequency: $\alpha(\theta_c, \Omega_P) = 1$. For angles greater than the critical coupling angle $\theta_c$, the plasmon is overdamped: the absorption spectrum is radiatively broadened and the absorptivity peak decreases.

Our quantum model allows the description of incandescent emission induced by injecting an electrical current in the highly doped layer sustaining the Berreman mode[13,27]. The application of an in-plane current induces the Joule heating of the electron gas, and thus increases the temperature of the electronic reservoir $T_e$ with respect to that of the photonic one, $T_{ph}$. In this out of equilibrium situation, the incandescent emission process can be seen as a plasmon-mediated exchange between the electronic and the photonic reservoir. The number of output photons, calculated by considering that both baths are in an incoherent thermal input state, is equal to the product of the absorptivity times the thermal occupancy at temperature $T_e$ [24]. This is equivalent to Kirchhoff's law of thermal emission, stating the equivalence between the emissivity and the absorptivity at a given frequency and emission angle. The incandescent power $P(\theta, \Omega_P)$ emitted at the plasma energy $\hbar\Omega_P$ and angle θ is thus proportional to $\alpha(\theta,\omega)/(\exp(\frac{\hbar\Omega_P}{k_B T})-1)$. In this configuration by modulating the injected current we can modulate the temperature of the electron gas and therefore the incandescent emission. We will show in the following that a modulation frequency as high as 50MHz has been measured, due to the small heat capacity of the electron gas.

# 4 High frequency incandescent emission

Figure 3(a) presents a sketch of our device, which is fabricated exploiting a highly doped 45nm GaInAs layer embedded between two undoped AlInAs barriers. This highly doped layer sustains a Berreman mode at 166meV, which is inserted in a microcavity of total thickness 1.125µm, delimited on the bottom by a highly doped 2µm thick GaInAs mirror. The sample has been grown by metal organic chemical vapor deposition and processed into a field effect transistor-like structure, consisting of two ohmic contacts made of TiAu. The gate size is L x W = 50 x 50 µm$^2$, where

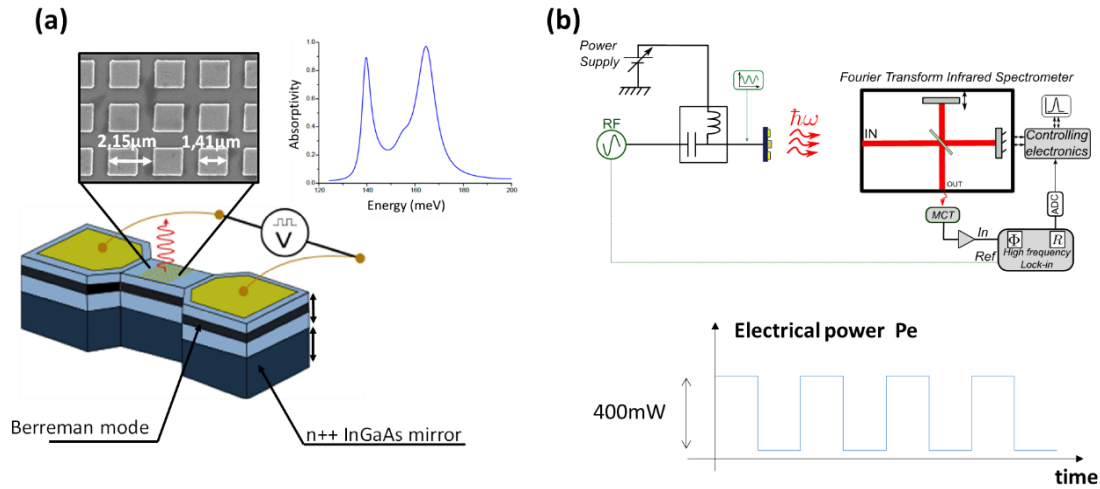

*Figure 3: (a) Sketch of the device fabricated for thermal emission experiments. It consists of a 50µm x 50µm mesa with a top grating for the extraction of incandescent emission. In order to electrically excite the thermal emission, two lateral Ti/Au contacts (125x100 µm$^2$) have been evaporated on the GaInAs layer sustaining the Berreman mode, after ICP etching. The top part of the panel presents a SEM image of the grating, with its characteristic dimensions, and the calculated absorption spectrum, displaying two polariton modes with almost unity peak absorptivity. (b) Sketch of the set-up used for high frequency modulation of thermal emission. An electrical current is injected between the two lateral contacts. Its time modulation results in a modulation of the emitted power.*

L and W stand for the length and width of the channel, respectively. On the top of the gate we have realized a two-dimensional metallic grating by e-beam lithography and Au evaporation. The top part of Fig. 3a presents a scanning electron microscope image of the grating, on which we have also indicated the dimensions of the square gold patches. The thickness of the cavity, defined as the distance between the bottom InGaAs mirror and the top grating, determines the energy of the cavity modes, while the ratio between the size of the patches and their periodicity determines the mode contrast, i.e. their coupling with free space. The right top part of Fig. 3a presents the calculated absorption spectrum of the device. Due to the huge dipole of the collective electronic excitation, the cavity mode strongly couples with the Berreman mode and gives rise to polariton modes, as indicated in the spectrum. The grating parameters and the cavity thickness have been designed in order to have two polaritonic modes with peak absorptivity, and hence peak emissivity, close to 1. Fig. 3b (top) shows the setup that have been used for our high frequency spectral resolved measurements. The plasmonic emitter is mounted in a high frequency holder to inject microwave into the electron gas that can also be biased with a dc power supplier. The emitted power is collected into a Fourier transform infrared spectrometer and then measured using a fast MCT detector with a frequency cutoff at 50 MHz. The spectra are measured in step-scan mode using a high frequency lock-in amplifier. In the bottom part of Fig. 3b we indicate the temporal form of the input current used to modulate the temperature of the electron gas.

Experimental results of the frequency modulated thermal emission are shown in Figure 4a (dotted curve). In the inset of the figure we provide spectra of the thermal emission for two frequency regimes. The spectra remain essentially identical, which proves that the emission signal arises from the incandescence of the polariton modes.

In order to model the frequency dependence of the emitted power shown in Figure 4a we developed a model that provides the temporal evolution of the electronic temperature $T_e$. The model is one-dimensional and assumes that the lattice temperature $T(z)$ depends on a single spatial variable $z$ (Figure 4b). The electronic temperature is different from the lattice temperature $T(z)$ and the thermal flux density that describes the heat exchanged between the electrons and the lattice is assumed to be of a simple form $\beta/L_{QW}(T_e - T(z=0))$. Here $L_{QW}$ is the quantum well thickness and $\beta$ is a coefficient that has the dimensions of a thermal conductivity [W/mK], and is treated as a fitting parameter (see further). Heat is generated only in the quantum well layer, through the dissipation of electrical power, and then propagates towards the heat sink at the bottom of the semiconductor substrate maintained at a temperature $T_0$ = 300K. In Figure 4b we indicate the different semiconductor layers between the quantum well and the heat sink.

Using energy conservation, we can write the following equations that couple the electronic and lattice temperature at the position of the quantum well $z = 0$:

$$c_e \frac{\partial T_e}{\partial t} = -\frac{\beta}{L_{QW}^2}(T_e - T(z=0)) + \frac{P_e(t)}{\sigma L_{QW}} \tag{6}$$

$$-\lambda_1 \left[\frac{\partial T}{\partial z}\right]_{z=0} = \frac{\beta}{L_{QW}}(T_e - T(z=0)) \tag{7}$$

with $c_e$ the thermal capacity of electrons, which can be assumed to be $c_e \approx 10^3$ J/m³K at 300K[28], $P_e(t)$ the electrical power dissipated in the quantum well of thickness $L_{QW}$ and surface $\sigma$ = 50x50 µm², and $\lambda_1$ = 5 W/mK the thermal conductivity of the InGaAs layer below the quantum well (Figure 4b). The lattice temperature $T(t,z)$ satisfies the Fourier heat equation in each layer of the structure:

$$\frac{\partial T}{\partial t} = D_i \frac{\partial^2 T}{\partial z^2} \tag{8}$$

Here $D_i$ are the thermal diffusion coefficients of the InGaAs ($i$ = 1) and InP ($i$ = 2) layers respectively. The temperature and the thermal flux are continuous at the InGaAs/InP interface. The boundary condition at the heat sink is $T(t, z = L_1 + L_2) = T_0$.

To analyze this problem $P_e(t)$ can be decomposed into Fourier series; then it is sufficient to examine the temperature response to a thermal source that is of the form $P_e(t) = <P_e> + P_1 \cos(\omega t)$. In the following, we will consider only the time-varying part of the temperature that is due to the source term $P_1 \cos(\omega t)$. The stationary part $<P_e>$ provides a static temperature response that is not relevant for the frequency dependence of the emitted power. In the experiment the electrical power is modulated using a square wave with a typical amplitude $P_0$ = 400mW (Figure 3b). The signal is detected using a lock-in amplifier, locked on the fundamental harmonic of the square wave and therefore in eq. (6) the time-varying source can be considered of the form $\left(\frac{2}{\pi}\right) P_0 \cos(\omega t)$. The temperature response thus splits into a static and a modulated contribution using a complex notation, $T = T_{st} + \Delta T(\omega)e^{i\omega t}$. By solving the above system with the appropriate boundary conditions we obtain the following expression for the modulated part of the complex electronic temperature:

$$\Delta T_e(\omega) = \frac{\Delta T_{e0}}{\frac{i\omega}{\omega_0} + \frac{1}{1+\frac{\beta L_1}{\lambda_1 L_{QW}}S(\omega)}}, \quad S(\omega) = \frac{1}{k_1 L_1} \frac{\tanh(2k_1L_1) + \frac{\lambda_1 k_1}{\lambda_2 k_2}\tanh(2k_2L_2)}{1 + \frac{\lambda_1 k_1}{\lambda_2 k_2}\tanh(2k_1L_1)\tanh(2k_2L_2)} \tag{9}$$

Here $\Delta T_{e0} = \left(\frac{2P_0}{\pi}\right)(L_{QW}/(\beta\sigma))$ and $\omega_0 = \beta/c_e L_{QW}^2$ is a cut-off frequency that describes the thermal dynamics of the electron gas alone. The frequency dependent coefficient $S(\omega)$ takes into account the temperature distribution in the InGaAs and InP layers. The quantities $k_i$ are defined as $k_i = (1+i)\sqrt{\omega/2D_i}$ and therefore, their reciprocal values, $1/k_i$, have the meaning of frequency dependent thermal diffusion lengths across the different layers.

Once the temperature $\Delta T_e(\omega)$ is known, the optical power can be deduced from the Bose-Einstein occupation factor of the plasmons modes $n_B(\hbar\Omega_P, T) = 1/(\exp\left(\frac{\hbar\Omega_P}{k_B T}\right) - 1)$ [13]: $P_{opt}(t) \propto n_B(\hbar\Omega_P, T_{st} + Re\{\Delta T_e(\omega)\exp(i\omega t)\})$ where $\hbar\Omega_P =$ 155 meV is the energy of the plasmon. The signal amplitude detected by the lock-in amplifier by our set-up described in Fig. 3b can therefore be modeled as:

$$P_{signal}(\omega) = P_0[n_B(\hbar\Omega_P, T_{st} + |\Delta T_e(\omega)|) - n_B(\hbar\Omega_P, T_{st} - |\Delta T_e(\omega)|)] \qquad (10)$$

$P_0$ is a constant adjusted so that the theoretical curve $P_{signal}(\omega)$ can be directly compared to the experimental data (Figure 4a).

All the physical parameters of the model are known except the constant $\beta$. The material parameters for the InGaAs are: $\lambda_1 =$ 5W/mK and $D_1 =$ 3.33 x10$^{-6}$ m²/s (the effect of the doping in the 2µm thick part has been neglected); while for InP: $\lambda_2 =$ 68W/mK and $D_2 =$ 41x10$^{-6}$ m²/s. The parameter $\beta$ is determined from the best fit of the experimental data which provides $\beta =$ 0.7 W/mK. As shown in Figure 4a a very good fit of the optical power is obtained in the entire frequency range, which covers almost 6 decades. In Figure 4a the bump observed at ~10$^5$ Hz is due to the InGaAs/InP interface. Therefore, up to 1 MHz the thermal dynamics of the system is dominated by the substrate underlying the electron gas. At higher frequencies the curve flattens as only the electronic dynamics is sufficiently fast to provide a temperature change. Indeed, from the value of $\beta =$ 0.7 W/mK we determine the thermal cut-off frequency of the electron gas: $\omega_0/(2\pi) =$ 55 GHz. Thermal sources based on electronic heating enable therefore an ultra-fast modulation of thermal radiation.

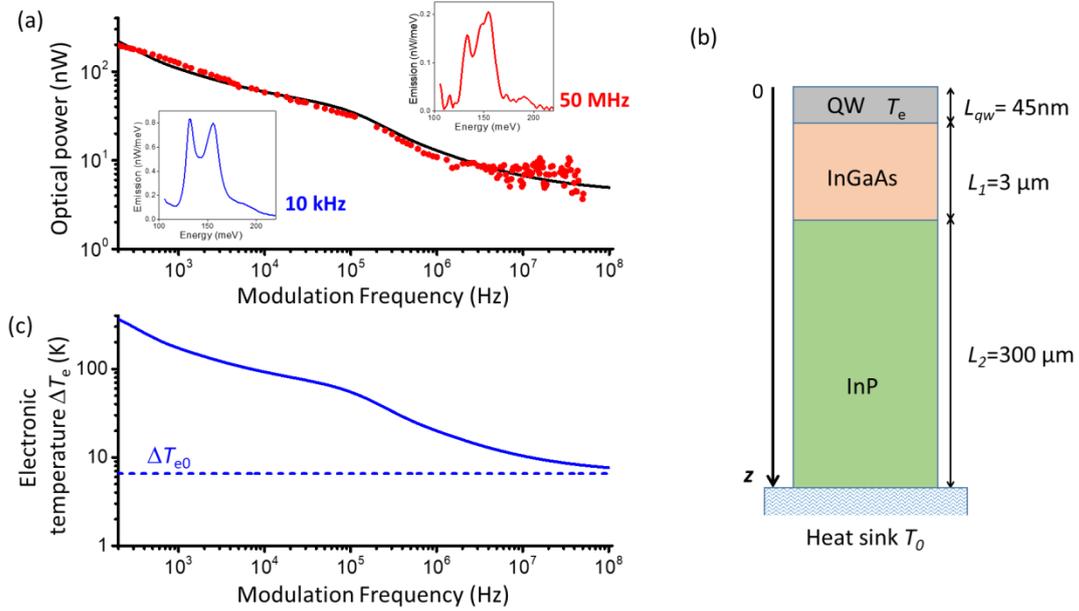

Figure 4: (a) Emitted power as a function of the modulation frequency (dots). The solid line is a result of modelling as described in the main text. The insets show two typical spectra respectively for modulation frequency of 10 kHz and 50 MHz. (b) Schematic of semiconductor layers from the quantum well down to the heat-sink used for our thermal model. (c) Estimations of the frequency-modulated part of the excess electronic temperature $\Delta T_e(\omega)$ resulting from eq.(9). The dashed line is the result of a simplified model where the effect of the semiconductor substrate has been neglected.

In Figure 4c, we plot the electronic temperature $|\Delta T_e(\omega)|$ predicted from eq. (9). Typical value observed at 10kHz is 100K, which corresponds to a thermal resistance $R_{th} = |\Delta T_e(\omega = 2\pi\, 10^4)|/P_0$ ~250 K/W, a value consistent with previous experiments[27]. Furthermore, the values of the excess electronic temperature are comparable with those reported in the literature for the case of quantum cascade lasers below threshold[29]. Note that in eq.(9), the effect of the substrate on the temperature dynamics of electrons is described by the function $S(\omega)$. It can be shown that this function vanishes at very high frequencies: $S(\omega \to \infty) \to 0$. In this limit the temperature of the substrate can be considered constant and does not affect the electron dynamics. eq. (9) becomes a low–pass filter, $\Delta T_e(\omega) = \Delta T_{e0}/(1 + i\omega/\omega_0)$ making evident the cut-off $\omega_0 = \beta/c_e L_{QW}^2$. In the current frequency range of Fig. 4c this simplified form appears as the flat part of the filter with $\Delta T_{e0}$ = 6.5 K. This model confirms that the effects of the substrate become negligible in the MHz range and the emission observed at 50 MHz, is already essentially due to the heating of the electron gas alone.

## 5 Conclusions and perspectives

We have exploited the peculiar quantum properties of plasmons confined in semiconductor layers to demonstrate perfect absorption and dynamic modulation of thermal emission up to 50 MHz in the mid-infrared frequency range. When they are confined in a layer with thickness smaller than the plasma wavelength, confined plasmons are optically active and display absorption resonances, known as Berreman modes, with a quality factor in the order of 20 at low internal angle of light propagation. Plasmon radiative rate increases as a function of the angle of emission and largely exceeds the non-radiative rate, a demonstration of the superradiant character of this collective electronic excitation. We have demonstrated that when the radiative rate equals the non-radiative one, a critical coupling regime occurs, corresponding to perfect absorption of the plasmon mode. All our experimental results are supported by a quantum model of the plasmon optical properties, which goes beyond Markov and rotating wave approximation.

The optical properties of the plasmon have been exploited to realize a perfect blackbody, in which incandescence is induced by Joule heating of the electron gas. We have shown that a time modulation of the injected current results in a modulation of the incandescent emission, up to 50 MHz, limited by the detector bandwidth. Our experimental results have been analyzed through an analytical model which takes into account the different dynamic behavior of electronic and lattice temperature under Joule heating. This model gives an excellent agreement with the experimental results in the entire frequency range, which covers almost 6 decades.

Berreman modes in semiconductors offer a huge possibility to control emissivity, wavelength, directionality and dynamic behavior of mid-infrared thermal sources in technologically mature material platforms. We believe that the versatility of this system can be beneficial to different field of applications of thermal emitters, particularly whenever a high frequency response is necessary, like for thermal camouflage, friend-or-foe identification or trace detection. Finally, the superradiant character of these excitations can be exploited also for radiative cooling in ultra-dense electronic circuits[30].

## 6 Acknowledgements

We acknowledge financial support from Agence Nationale de la Recherche (Grant ANR-14-CE26-0023-01). We thank Stephan Suffit for help with the device fabrication, E. Maggiolini for preliminary experiments on high modulation of thermal emission, J. J. Greffet for the loan of the 50 MHz-bandwidth MCT detector, E. Sakat for discussions on the cavity design.

# Bibliography

1. Ferrell, R. A. Predicted Radiation of Plasma Oscillations in Metal Films. *Phys. Rev.* **111**, 1214–1222 (1958).

2. Melnyk, A. R. & Harrison, M. J. Resonant Excitation of Plasmons in Thin Films by Elecromagnetic Waves. *Phys. Rev. Lett.* **21**, 85–88 (1968).

3. Lindau, I. & Nilsson, P. O. Experimental Verification of Optically Excited Longitudinal Plasmons. *Physica Scripta* **3**, 87–92 (1971).

4. Melnyk, A. R. & Harrison, M. J. Theory of Optical Excitation of Plasmons in Metals. *Phys. Rev. B* **2**, 835–850 (1970).

5. Scholl, J. A., Koh, A. L. & Dionne, J. A. Quantum plasmon resonances of individual metallic nanoparticles. *Nature* **483**, 421–427 (2012).

6. Halperin, W. P. Quantum size effects in metal particles. *Rev. Mod. Phys.* **58**, 533–606 (1986).

7. Mühlschlegel, P., Eisler, H.-J., Martin, O. J. F., Hecht, B. & Pohl, D. W. Resonant Optical Antennas. *Science* **308**, 1607 (2005).

8. Schuck, P. J., Fromm, D. P., Sundaramurthy, A., Kino, G. S. & Moerner, W. E. Improving the Mismatch between Light and Nanoscale Objects with Gold Bowtie Nanoantennas. *Phys. Rev. Lett.* **94**, 017402 (2005).

9. Tame, M. S. *et al.* Quantum plasmonics. *Nature Physics* **9**, 329–340 (2013).

10. Mason, J. A., Smith, S. & Wasserman, D. Strong absorption and selective thermal emission from a midinfrared metamaterial. *Appl. Phys. Lett.* **98**, 241105 (2011).

11. Huppert, S. *et al.* Radiatively Broadened Incandescent Sources. *ACS Photonics* **2**, 1663–1668 (2015).

12. Campione, S. *et al.* Directional and monochromatic thermal emitter from epsilon-near-zero conditions in semiconductor hyperbolic metamaterials. *Scientific Reports* **6**, 34746 (2016).

13. Laurent, T. *et al.* Superradiant Emission from a Collective Excitation in a Semiconductor. *Phys. Rev. Lett.* **115**, 187402 (2015).

14. Baranov, D. G. *et al.* Nanophotonic engineering of far-field thermal emitters. *Nature Materials* **18**, 920–930 (2019).

15. Greffet, J.-J. *et al.* Coherent emission of light by thermal sources. *Nature* **416**, 61–64 (2002).

16. Lu, G. *et al.* Narrowband Polaritonic Thermal Emitters Driven by Waste Heat. *ACS Omega* **5**, 10900–10908 (2020).

17. Wang, T. *et al.* Phonon-Polaritonic Bowtie Nanoantennas: Controlling Infrared Thermal Radiation at the Nanoscale. *ACS Photonics* **4**, 1753–1760 (2017).

18. De Zoysa, M. *et al.* Conversion of broadband to narrowband thermal emission through energy recycling. *Nature Photonics* **6**, 535–539 (2012).

19. Yang, Z.-Y. *et al.* Narrowband Wavelength Selective Thermal Emitters by Confined Tamm Plasmon Polaritons. *ACS Photonics* **4**, 2212–2219 (2017).

20. Inoue, T., Zoysa, M. D., Asano, T. & Noda, S. Realization of dynamic thermal emission control. *Nature Materials* **13**, 928–931 (2014).

21. Askenazi, B. *et al.* Ultra-strong light–matter coupling for designer Reststrahlen band. *New Journal of Physics* **16**, 043029 (2014).

22. Vasanelli, A. *et al.* Semiconductor quantum plasmonics.

23. Pegolotti, G., Vasanelli, A., Todorov, Y. & Sirtori, C. Quantum model of coupled intersubband plasmons. *Physical Review B* **90**, (2014).

24. Huppert, S., Vasanelli, A., Pegolotti, G., Todorov, Y. & Sirtori, C. Strong and ultrastrong coupling with free-space radiation. *Physical Review B* **94**, (2016).

25. Frucci, G. *et al.* Cooperative Lamb shift and superradiance in an optoelectronic device. *New Journal of Physics* **19**, 043006 (2017).

26. Ciuti, C., Bastard, G. & Carusotto, I. Quantum vacuum properties of the intersubband cavity polariton field. *Phys. Rev. B* **72**, 115303 (2005).

27. Laurent, T. *et al.* Electrical excitation of superradiant intersubband plasmons. *Appl. Phys. Lett.* **107**, 241112 (2015).

28. Chakravarti, A. N., Ghatak, K. P., Ghosh, S. & Chowdhury, A. K. Effect of Heavy Doping on the Electronic Heat Capacity in Semiconductors. *physica status solidi (b)* **109**, 705–710 (1982).

29. Spagnolo, V., Scamarcio, G., Page, H. & Sirtori, C. Simultaneous measurement of the electronic and lattice temperatures in GaAs/Al0.45Ga0.55As quantum-cascade lasers: Influence on the optical performance. *Appl. Phys. Lett.* **84**, 3690–3692 (2004).

30. Buddhiraju, S., Li, W. & Fan, S. Photonic Refrigeration from Time-Modulated Thermal Emission. *Phys. Rev. Lett.* **124**, 077402 (2020).